\documentclass[12pt]{article}

\usepackage[totalheight = 23cm, totalwidth = 17cm]{geometry}
\usepackage{amssymb,amsfonts,amsmath,graphicx}
\usepackage{enumerate}

\usepackage{color}

\newcommand{\eq}[1]{Eq.~(\ref{#1})}
\newcommand{\mpl}{m_{\rm Pl}}
\newcommand{\calA}{{\cal A}}
\newcommand{\calD}{{\cal D}}
\newcommand{\calI}{{\cal I}}
\newcommand{\calO}{{\cal O}}
\newcommand{\calP}{{\cal P}}
\newcommand{\calR}{{\cal R}}
\newcommand{\calW}{{\cal W}}

\begin{document}

\begin{titlepage}

\rightline{\footnotesize{APCTP-Pre2021-025}} 
\vspace{-0.2cm}

\begin{center}

\vskip 1.0cm

\Large{\bf 
Potential reconstruction from general power spectrum \\ in single-field inflation
}

\vskip 1.0cm

\normalsize{
Ki-Young Choi$^{a}$,
Jinn-Ouk Gong$^{b,c}$,
Su-beom Kang$^{a}$
and
Rathul Nath Raveendran$^{a}$
}

\vskip 0.5cm

{\it
$^{a}$Department of Physics, Sungkyunkwan University, Suwon 16419, Korea
\\
$^{b}$Department of Science Education, Ewha Womans University, Seoul 03760, Korea
\\
$^{c}$Asia Pacific Center for Theoretical Physics, Pohang 37673, Korea
}

\vskip 1.2cm

\end{center}

\begin{abstract}

We suggest a new method to reconstruct, within canonical single-field inflation, the inflaton potential directly from the primordial power spectrum which may deviate significantly from near scale-invariance. Our approach relies on a more generalized slow-roll approximation than the standard one, and can probe the properties of the inflaton potential reliably. We give a few examples for reconstructing potential and discuss the validity of our method.

\end{abstract}

\end{titlepage}

\setcounter{page}{0}
\newpage
\setcounter{page}{1}

\section{Introduction}
\label{sec:intro}


The recent observations on the temperature anisotropies of the cosmic microwave background (CMB) by the Planck mission~\cite{Aghanim:2018eyx} and the large-scale distribution of galaxies by spectroscopic and photometric surveys~\cite{eBOSS:2020yzd,DES:2021wwk} have confirmed important properties of the primordial seed perturbations. These perturbations, from the onset of the hot big bang evolution of the universe, are adiabatic, have almost scale-invariant power spectrum, and follow nearly perfect Gaussian statistics. Remarkably, these properties of the primordial perturbations are naturally satisfied within the framework of cosmic inflation~\cite{Guth:1980zm,Linde:1981mu,Albrecht:1982wi} -- a phase of accelerated expansion of the universe. The standard picture of the generation of the primordial perturbations is as follows~\cite{Mukhanov:2005sc,Weinberg:2008zzc,Lyth:2009zz}. During inflation, the vacuum quantum fluctuations on small scales are stretched beyond the horizon scale. On super-horizon scales, these fluctuations become classical and manifest themselves as the curvature perturbation on the constant-time hypersurfaces. After inflation, such curvature perturbation leads to the inhomogeneities in matter distribution, which are amplified by gravitational instability and become observable structure in the universe like galaxies and clusters of galaxies. The curvature perturbation, as the seed of the observed CMB anisotropies and inhomogeneous distribution of galaxies on large-scales, has the very desired properties constrained by precise cosmological observations~\cite{Planck:2018jri}.


To implement a phase of inflation, we need a special matter content that possesses a negative pressure. We typically invoke a hypothetical scalar field, called the inflaton, with a sufficient flat potential that dominates the universe during inflation. The motion of the inflaton is dragged by the Hubble friction, so that it rolls slowly down the potential. The universe experiences a quasi-exponential expansion until this ``slow-roll'' period comes to an end. Thus, under the picture of slow-roll inflation, we can resort to the form of the potential to calculate the power spectrum of the curvature perturbation that can be constrained directly by observations. Usually, one assumes an ad-hoc potential, possibly motivated by theories of high energy physics, to construct a model of inflation~\cite{Lyth:1998xn,Mazumdar:2010sa}. Then the viability of the model is examined by calculating the number of $e$-folds, the power spectra of the scalar and tensor perturbations, as well as possible non-Gaussianities to compare with the observational data. Under the slow-roll approximation, these calculations are straightforward and the predictions depend on the functional form of the potential. The zoology of the viable models of slow-roll inflation is vast~\cite{Martin:2013tda}.


However, there is no a priori reason why one model is more favourable than others. Exactly speaking, certain models may enjoy theoretical advantages: they have less free parameters, the couplings of the models are less finely tuned, they can be embedded in the minimal extension of the standard model of particle physics, and so on. Such models would be theoretically more aesthetic than others. But regarding observational viability, as long as observational constraints are satisfied within the same confidence level, every model is equivalent. Therefore, an attractive alternative to the model building of slow-roll inflation is, directly from the given power spectrum, to reconstruct the potential. This reconstruction programme was first pioneered in~\cite{Hodges:1990bf}, where the inflaton potential was presented as an integral of the given scalar power spectrum, assuming that inflation is dominated by a single inflaton field using the standard slow-roll approximation. The reconstructed potential in such a way is determined up to an unknown constant, which leads to different functional forms. This constant can be fixed by generalizing this approach to include the tensor modes~\cite{Copeland:1993jj,Copeland:1993ie}: observational constraints on the tensor spectrum at a single scale can fix the undetermined constant, and can allow us to determine the potential uniquely~\cite{Lidsey:1995np}. The full knowledge of the tensor power spectrum can provide a consistency check compared with the scalar power spectrum. The reconstruction programme has received renewed interests recently in the context of the primordial black holes (PBHs) and the power spectrum that allows their formation~\cite{Hertzberg:2017dkh,Choi:2021yxz}.


In reconstructing the inflaton potential from the power spectrum, the slow-roll approximation is the key machinery. This is reasonable because the number of $e$-folds elapsed on the CMB scales constrained by observations is $\calO(5-10)$, well before the end of inflation, so that the slow-roll motion of the inflaton is very likely to be effective. However, the {\it additional} assumption that the so-called slow-roll parameters are approximately constant is not justified: we may only keep the assumption that both the Hubble parameter and the velocity of the inflaton during inflation vary slowly, which is fully consistent with the current observational constraints. Furthermore, it is not guaranteed at all that the slow-roll approximation remains valid beyond the CMB scales. If not, there should exist deviations from the featureless, slightly red-tilted power spectrum as preferred on the CMB scales -- such as bumps and oscillations that lead to interesting observational consequences like the formation of the PBHs. Thus, generalizing the standard slow-roll approximation for the reconstruction programme encompasses broad classes of the power spectrum with rich phenomenologies, enhancing greatly our accessibility to the inflationary dynamics.

In this article, we adopt the general slow-roll approximation (GSR)~\cite{Dodelson:2001sh,Stewart:2001cd,Choe:2004zg,Gong:2004kd} and show a method for  reconstructing the inflaton potential directly from a given power spectrum. This article is outlined as follows. In Section~\ref{sec:reconstruction} we review GSR and introduce our method for reconstructing the inflaton potential. In Section~\ref{sec:examples} we show a few examples using our method, and conclude shortly in Section~\ref{sec:conc}.

\section{GSR and potential reconstruction}
\label{sec:reconstruction}

The power spectrum $\calP_\calR$ of the comoving curvature perturbation $\calR$, constrained from the CMB anisotropies, exhibits a small deviation from the flat one. This indicates that during inflation, the slow-roll parameters 
\begin{align}
\label{eq:sr1}
\epsilon & \equiv - \frac{\dot{H}}{H^2} = -\frac{d\log{H}}{d\log{a}} \, ,
\\
\label{eq:sr2}
\delta_1 & \equiv \frac{\ddot\phi}{H\dot\phi} = \frac{d\log\dot\phi}{d\log{a}} \, ,
\end{align}
are small, say, $\calO(\varepsilon)$ for some small parameter $\varepsilon$. In the standard slow-roll approximation, however, it is additionally assumed these parameters are also nearly scale-invariant so they vary at $\calO(\varepsilon^2)$. In GSR, we abandon this extra assumption and consider all the deviations from the perfect de Sitter expansion on equal footing.

For each Fourier mode of the comoving curvature perturbation $\calR$, defining
\begin{align}
\varphi & \equiv z\calR \, ,
\\
z & \equiv \frac{a\dot\phi}{H} \, ,
\end{align}
then the following equation is satisfied:
\begin{equation}
\label{eq:modeeq1}
\frac{d^2\varphi}{d\xi^2} + \bigg( k^2 - \frac{1}{z} \frac{d^2z}{d\xi^2} \bigg) \varphi = 0 \, .
\end{equation}
Here, we have defined the positive conformal time $\xi$:
\begin{equation}
\label{eq:xi-def}
\xi \equiv - \int \frac{dt}{a} = \frac{1}{aH} \big[ 1 + \calO(\epsilon) \big] \, .
\end{equation}

Now, we can rewrite \eq{eq:modeeq1} so that the contribution for the exactly scale-invariant power spectrum and that for the departure from such a spectrum are manifest.
By defining $y \equiv \sqrt{2k}\varphi$ and $x \equiv k\xi$ and rescaling $z$ as
\begin{equation}
\label{eq:GSRfunction}
f(\log\xi) \equiv \frac{2\pi x}{k}z = 2\pi\xi \frac{a\dot\phi}{H} \, ,
\end{equation}
\eq{eq:modeeq1} becomes
\begin{equation}
\label{eq:modeeq2}
\frac{d^2y}{dx^2} + \bigg( 1 - \frac{2}{x^2} \bigg) y = \frac{1}{x^2} g(\log\xi) y \, ,
\end{equation}
where
\begin{equation}
g \equiv \frac{f''-3f'}{f},
\end{equation}
with $f' \equiv df/d\log\xi$. From \eq{eq:xi-def}, we can note that $f$ is to the zeroth order in slow-roll the power spectrum, $f \approx \sqrt{\calP_\calR}$ (see below). Thus we can separate systematically the contributions suppressed and not suppressed by the slow-roll parameters from the beginning. In \eq{eq:modeeq2}, the left-hand side represents the mode function equation in the perfect de Sitter background and the corresponding power spectrum of the solution, conveniently written as
\begin{equation}
\label{eq:GSRpower}
\calP_\calR(k) = \lim_{x\to 0} \bigg| \frac{xy}{f} \bigg|^2 \, ,
\end{equation}
is exactly scale-invariant. The function $g$ on the right-hand side of \eq{eq:modeeq2} represents, as mentioned above, all the possible deviations from the scale-invariance of the power spectrum. It includes the contributions suppressed by the slow-roll parameters Eqs.~\eqref{eq:sr1} and \eqref{eq:sr2} and their time variations, which are all equivalent here. Thus the expansion parameter of GSR is the function $g$ itself.

To the leading order of GSR, i.e. up to $\calO(g)$, we can solve for $y(x)$ using the Green's function method~\cite{Gong:2001he,Gong:2002cx} and from \eq{eq:GSRpower} the power spectrum can be expressed as~\cite{Stewart:2001cd,Choe:2004zg}
\begin{equation}
\label{eq:GSRspectrum}
\log\calP_\calR(k)
=
\int_0^\infty \frac{d\xi}{\xi} \big[ -k\xi W'(k\xi) \big]
\bigg[ \log \bigg( \frac{1}{f^2} \bigg) + \frac{2}{3} \frac{f'}{f} + \calO(g^2) \bigg]
\, ,
\end{equation}
where $W(x)$ is the following window function:
\begin{equation}
\label{eq:window}
W(x) = \frac{3\sin(2x)}{2x^3} - \frac{3\cos(2x)}{x^2} - \frac{3\sin(2x)}{2x} - 1 \, .
\end{equation}
%
In the standard slow-roll approximation, as can be read from \eq{eq:GSRspectrum}, $f'$ is slowly-varying and the power spectrum is to the zeroth order given by the well-known result $\calP_\calR \approx [H^2/(2\pi \dot{\phi})]^2$. This relation was used to reconstruct the potential in the previous literature on the reconstruction programme~\cite{Hodges:1990bf,Copeland:1993jj,Copeland:1993ie}. However in GSR, $f'/f$ may vary rapidly while it remains small, and thus gives notable features in the power spectrum. Furthermore, in GSR the evolution of each mode throughout the entire inflationary epoch is taken into account automatically as the range of integration shows, while in the standard slow-roll approximation one only considers the effects around the moment of horizon crossing, $k=aH$.

Mathematically, \eq{eq:GSRspectrum} represents an integral transformation between $\calP_\calR(k)$ and $f(\log\xi)$ via the window function \eq{eq:window}. Thus using the convolution theorem we can write 
\begin{equation}
\label{eq:inverse}
\log \bigg( \frac{1}{f^2} \bigg) = \int_0^\infty \frac{dk}{k} m(k\xi)\log\calP_\calR(k) 
\end{equation}
for some function $m$. Substituting this expression into \eq{eq:GSRspectrum} and explicitly calculating the integral, and taking account of convergence as the argument of $m$ goes to infinity, we can find a formal inverse function valid to the leading order in GSR~\cite{Joy:2005ep,Joy:2005pe}:
\begin{equation}
m(x) = \frac{2}{\pi} \bigg[ \frac{1}{x} - \frac{\cos(2x)}{x} - \sin(2x) \bigg] \, .
\end{equation}

Now, assuming that the matter sector is canonical, we can find the following equation for the Hubble parameter:
\begin{equation}
\label{eq:Heq}
\dot{H} = - \frac{\dot\phi^2}{2\mpl^2} = - \frac{H^2f^2}{2(2\pi)^2\mpl^2 a^2\xi^2} \, ,
\end{equation}
where in the second equality we have used \eq{eq:GSRfunction} to eliminate $\dot\phi$. In the slow-roll limit $\epsilon \ll 1$, using $\xi \approx 1/(aH)$, we find the following differential equation for $H$:
\begin{equation}
H^{-3} \frac{dH}{d\xi} = \frac{1}{2(2\pi)^2\mpl^2} \frac{f^2}{\xi} \, ,
\end{equation}
where the unknown function $f(\log\xi)$ is determined from the given power spectrum by \eq{eq:inverse}. The integration gives $H$ in terms of $\xi$ as
\begin{equation}
\label{eq:Hubble}
\frac{1}{H^2} = \frac{1}{H_i^2} - \frac{1}{(2\pi)^2\mpl^2} \int_{\xi_i}^\xi f^2(\log\xi') d\log\xi' \, .
\end{equation}
Note that if the second term is small compared to the integration constant $1/H_i^2$, we can approximate $H \approx H_i$ to leading order in GSR. Meanwhile, from the Friedmann equation with \eq{eq:GSRfunction}, we can write easily the potential as a function of $\log\xi$:
\begin{equation}
\label{eq:potential}
V = 3\mpl^2H^2 - \frac{1}{2}\dot\phi^2
= 3\mpl^2H^2 \bigg[ 1 - \frac{f^2H^2}{6(2\pi)^2\mpl^2} \bigg]
\, .
\end{equation}
Finally, from \eq{eq:GSRfunction} we find the differential equation for $\phi$:
\begin{equation}
\label{eq:phieq}
\frac{d\phi}{d\log\xi} = -\frac{fH}{2\pi} \, ,
\end{equation}
and integrating this equation gives 
\begin{equation}
\label{eq:phi}
\phi = \phi_i - \int_{\xi_i}^\xi \frac{fH}{2\pi} d\log\xi \, .
\end{equation}
By eliminating $\xi$ in \eq{eq:potential} using \eq{eq:phi}, finally we can reconstruct the potential as a function of the inflaton field, $V(\phi)$. This is the main result of this article.

It is worthwhile to mention that Eqs.~\eqref{eq:Heq}, \eqref{eq:potential} and \eqref{eq:phieq} closely resemble the Hamilton-Jacobi approach to the inflationary dynamics~\cite{Kinney:1997ne}, in the sense that essentially $H$ is regarded as a function of the inflaton $\phi$, rather than time $\xi$. The critical difference from the standard Hamilton-Jacobi approach is that the time derivative of $\phi$ is not related to $dH/d\phi$ but to $H$ itself, as given in \eq{eq:phieq}. This allows, for a monotonic evolution of $\phi$, an exact one-to-one correspondence between $\xi$ and $\phi$ directly given the fundamental GSR function $f(\log\xi)$ inferred from \eq{eq:inverse}.

\section{Examples}
\label{sec:examples}

\subsection{Power-law spectrum}

First, we consider a power-law spectrum given by
\begin{equation}
\label{eq:power-law}
\calP_\calR(k) = A_s \bigg( \frac{k}{k_*} \bigg)^{n_s-1} \, .
\end{equation}
Then \eq{eq:inverse} gives
\begin{equation}
\log \bigg( \frac{1}{f^2} \bigg) = \log{A_s} + (n_s-1) \big[ \alpha - \log(k_*\xi) \big] \, ,
\end{equation}
where $\alpha \equiv 2 - \log2 - \gamma \approx 0.729637$, with $\gamma \approx 0.577216$ being the Euler-Mascheroni constant. That is,
\begin{equation}
\label{eq:ex1-f}
f^2 = \frac{(k_*\xi)^{n_s-1}}{A_s e^{\alpha(n_s-1)}} \, .
\end{equation}
Then \eq{eq:Hubble} gives
\begin{equation}
\frac{1}{H^2} = \frac{1}{H_i^2} \bigg[ 1 - \frac{(k_*\xi_i)^{n_s-1}}{\beta} + \frac{(k_*\xi)^{n_s-1}}{\beta} \bigg] \, ,
\end{equation}
where $H_i$ is the value of the Hubble parameter at $\xi = \xi_i$, and $\beta$ is a positive constant defined by
\begin{equation}
\beta \equiv 4\pi^2(1-n_s)A_s \frac{\mpl^2}{H_i^2} e^{\alpha(n_s-1)} \, .
\end{equation}
Using $H$ above along with $f$ in \eq{eq:phi}, we find
\begin{equation}
\label{eq:ex1-phi}
\Delta\phi \equiv \phi - \phi_i = - \frac{2\mpl}{\sqrt{1-n_s}} \sinh^{-1}
\Bigg[ \sqrt{\frac{(k_*\xi')^{n_s-1}}{\beta-(k_*\xi_i)^{n_s-1}}} \Bigg] \Bigg|_{\xi'=\xi_i}^{\xi'=\xi}
\, ,
\end{equation}
where we have chosen the minus sign for $f$ from \eq{eq:ex1-f} to make the field decreasing as inflation goes on. We can absorb the  $k_*\xi_i$ term into the constant $\phi_0$ as 
\begin{equation}
- \sinh^{-1} \Bigg[ \sqrt{\frac{(k_*\xi)^{n_s-1}}{\beta-(k_*\xi_i)^{n_s-1}}} \Bigg] 
= 
\Delta \phi  + \sinh^{-1} \Bigg[ \sqrt{\frac{(k_*\xi_i)^{n_s-1}}{\beta-(k_*\xi_i)^{n_s-1}}} \Bigg] 
\equiv 
\phi -\phi_0 
\, .
\end{equation}
Finally, the potential can be written in terms of $\phi$ from \eq{eq:potential} as
\begin{equation}
\label{eq:ex1-V}
V(\phi) = \frac{3\mpl^2H_i^2\beta}{\beta-(k_*\xi_i)^{n_s-1}}
\frac{ 1 - \frac16 (1-n_s) \tanh^2 \left[ \sqrt{1-n_s}\frac{\phi-\phi_0}{2\mpl}\right] }
{ 1 + \sinh^2 \left[ \sqrt{1-n_s}\frac{\phi-\phi_0}{2\mpl}\right] }
\, .
\end{equation}
Note that if we assume $H_i$ is the value of the Hubble parameter relevant for the CMB scales, we find $\beta \approx 0.282195/r$ using the central value $n_s = 0.9656$, where $r$ is the tensor-to-scalar ratio. Considering the current bound $r < 0.056$~\cite{Planck:2018jri}, we find that $\beta > (k_*\xi_i)^{n_s-1}$ for $k_*\xi_i > 3.82402 \times 10^{-21}$. This includes practically all the observable scales, once $\xi_i$ is chosen not too far from the CMB scales.

\begin{description}

\item [Small-field potential] 

First we consider $\beta \gg (k_*\xi_i)^{n_s-1}$. In this case, \eq{eq:ex1-phi} is approximated as
\begin{equation}
\Delta\phi \approx - \frac{2\mpl}{\sqrt{1-n_s}} \sqrt{\frac{(k_*\xi)^{n_s-1}}{\beta}} \, ,
\end{equation}
and \eq{eq:ex1-V} becomes the following simple form:
\begin{equation}
V(\phi) \approx 3\mpl^2H_i^2 \bigg[ 1 - \frac{1-n_s}{4} (\Delta\phi)^2 \bigg] \, .
\end{equation}
This potential is vacuum-dominated, with small field variations. This is reasonable as a small value of $r$ indicates a small value of $\epsilon$, which corresponds to a small field excursion.

\item [Large-field potential] 

The opposite limit, $\beta \ll (k_*\xi_i)^{n_s-1}$, is interesting -- not consistent with observations though, since this case corresponds to $r \gtrsim 0.32$. Then, $H^{-2} \approx H_i^{-2} (k_*\xi)^{n_s-1}/\beta$, so that $f^2H^2 \approx H_i\beta e^{\alpha(1-n_s)}/A_s$ is constant. Thus the field variation becomes
\begin{equation}
\Delta\phi \approx \sqrt{1-n_s} \mpl \log \bigg( \frac{\xi}{\xi_i} \bigg) \, ,
\end{equation}
so that we can absorb $k_*\xi_i$ as follows:
\begin{equation}
\log(k_*\xi) = \frac{\Delta\phi}{\sqrt{1-n_s}\mpl} + \log(k_*\xi_i) \equiv \frac{\phi-\phi_0}{\sqrt{1-n_s}\mpl} \, .
\end{equation}
Then, the reconstructed potential is obtained as an exponential function:
\begin{equation}
V(\phi) = 3\mpl^2H_i^2 \beta \exp \bigg( \sqrt{1-n_s} \frac{\phi-\phi_0}{\mpl} \bigg) \, .
\end{equation}

\end{description}

\subsection{Featured power spectrum}
\label{sec:FeaturedP}

A sufficiently flat and slowly varying inflaton potential gives a featureless and nearly scale-invariant power spectrum. However, the outliers of the CMB power spectrum may imply, though not decisively preferred as of now, occasional departures from the usual slow-roll phase. This can, with the matter sector being canonical, be achieved by an inflaton potential that possesses some features~\cite{Covi:2006ci} -- deviations from otherwise smooth potential. Since the standard slow-roll approximation is broken across the features in the potential, usually one resorts to the numerical approach, and/or to restricted analytic approximations such as matching the mode function solutions at the features. With GSR, however, we have a fully valid analytic machinery to deal with features. Especially, if the features -- if exist at all -- are small as indicated by the CMB observations, we may obtain a fully analytic formula for the power spectrum. In this section, we consider such a fortunate case -- the power spectrum with features is given analytically, from which we directly reconstruct the inflaton potential.

Since the possible analytic form of the power spectrum is diverse depending on the potential and features~\cite{Gong:2005jr}, we just consider as an example the following form for the localized oscillatory features~\cite{Planck:2018jri,Miranda:2013wxa} which results from a $\tanh$ step in the potential~\cite{Adams:2001vc}:
\begin{equation}
\label{eq:featuredP}
\log\calP_\calR(k) 
=
\log\calP_\calR^0(k) + \calI_0(k) + \log\big[ 1 + \calI_1^2(k) \big] \, ,
\end{equation}
where $\calP_\calR^0$ is the power-law power spectrum \eq{eq:power-law}, and the various functions are defined as follows:
\begin{equation}
\begin{split}
\calI_0(k)
& =
\calA_s \calW_0(k/k_s) \calD \bigg( \frac{k/k_s}{x_s} \bigg) 
\, ,
\\
\calI_1(k)
& =
\frac{1}{\sqrt{2}} \bigg[ \frac{\pi}{2} (1-n_s) + \calA_s \calW_1(k/k_s) \calD \bigg( \frac{k/k_s}{x_s} \bigg) \bigg] 
\, ,
\\
\calW_0(x)
& =
\frac{1}{2x^4} \big[ (18x-6x^3)\cos(2x) + (15x^2-9)\sin(2x) \big]
\, ,
\\
\calW_1(x)
& =
-\frac{3}{x^4} (x\cos{x}-\sin{x}) \big[ 3x\cos{x} + (2x^2-3)\sin{x} \big]
\, ,
\\
\calD(x) 
& =
\frac{x}{\sinh{x}}
\, .
\end{split}
\end{equation}
With the pivot scale for the power-law spectrum being fixed at $k_* = 5\times10^{-2}$ Mpc$^{-1}$, the best-fit parameters are as follows~\cite{Planck:2018jri}:
\begin{equation}
A_s = 2.0989 \times 10^{-9} \, ,
\quad
n_s = 0.9649 \, ,
\quad
\calA_s = 0.38 \, ,
\quad
\log_{10} k_x = -3.09 \, ,
\quad
\log{x}_s = 0.15 \, .
\end{equation}

\begin{figure}[t!]
\begin{center}
\includegraphics[width = 0.45\textwidth]{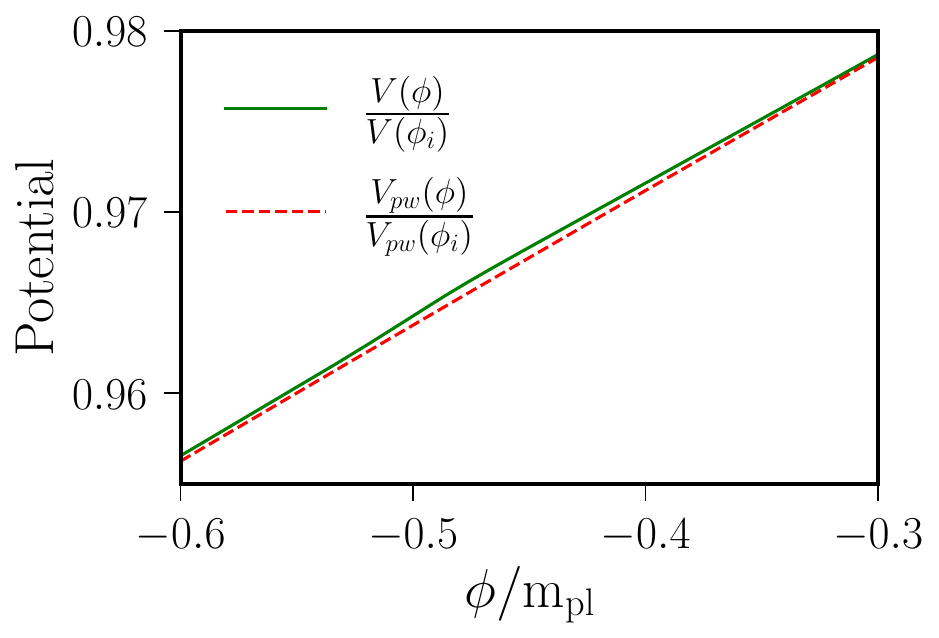}
\quad
\includegraphics[width = 0.44\textwidth]{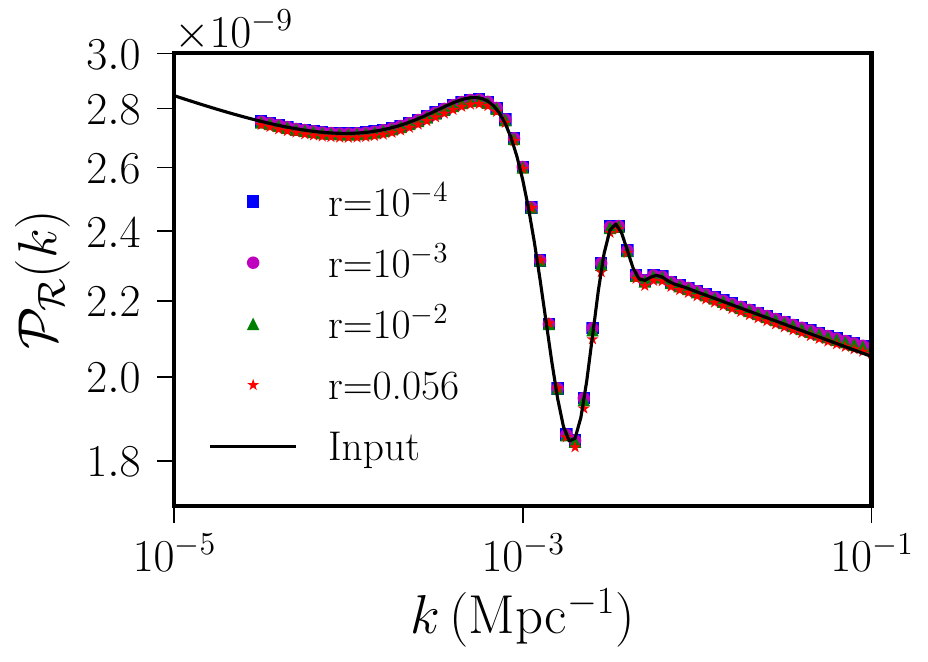}
\end{center}
\caption{(Left panel) Reconstructed potential from the featured power spectrum \eqref{eq:featuredP} for $r=0.056$ and $\phi_i=0$. The potential for the power-law power spectrum is shown with dashed line for comparison. (Right panel) Power spectra calculated numerically from the reconstructed potential for different values of $r$ as shown in the figure. The input power spectrum (solid line) is also shown for comparison.}
\label{fig:featuredP}
\end{figure}

In the left panel of Figure~\ref{fig:featuredP}, we show the reconstructed potential for the tensor-to-scalar ratio $r=0.056$ with the initial value $\phi_i=0$. Indeed, as we can see, there is a smooth, tiny step in the reconstructed potential. Since the height of the step is very tiny -- $\calO(0.1)$\% or even smaller change of the potential -- we can expect that GSR is very effective. Indeed, the magnitude of the function $g$, the expansion parameter of GSR [see the discussions below \eq{eq:GSRpower}], remains smaller than 1 throughout so that GSR expansion is fully justified. As a consistency check, in the right panel of Figure~\ref{fig:featuredP} we have calculated the power spectra from the reconstructed potential by exact numerical calculations for several different values of the tensor-to-scalar ratio $r=10^{-4}$, $10^{-3}$, $10^{-2}$, and $0.056$. The original power spectrum \eq{eq:featuredP}, which is the input to \eq{eq:inverse} for the reconstruction programe, is also shown for comparison with a solid line.

\subsection{Power spectrum with a peak}
\label{sec:PeakP}

As the final example, we consider the following power spectrum:
\begin{equation}
\label{eq:peakP}
\calP_\calR(k)
=
\calP_\calR^0(k) \Bigg\{ 1 + A_p \exp \left[ -\bigg( \frac{\log_{10}(k/k_c)}{\Delta} \bigg)^2 \right] \Bigg\} \, ,
\end{equation}
where $\calP_\calR^0$ is the power-law power spectrum \eq{eq:power-law}. Thus, this spectrum exhibits a Gaussian peak of height $A_p$ and width $\Delta$ centered at the scale $k=k_c$. This spectrum may well lead to the formation of PBHs if $\calP_\calR = \calO(0.01)$.

\begin{figure}[t!]
\begin{center}
\includegraphics[width = 0.48\textwidth]{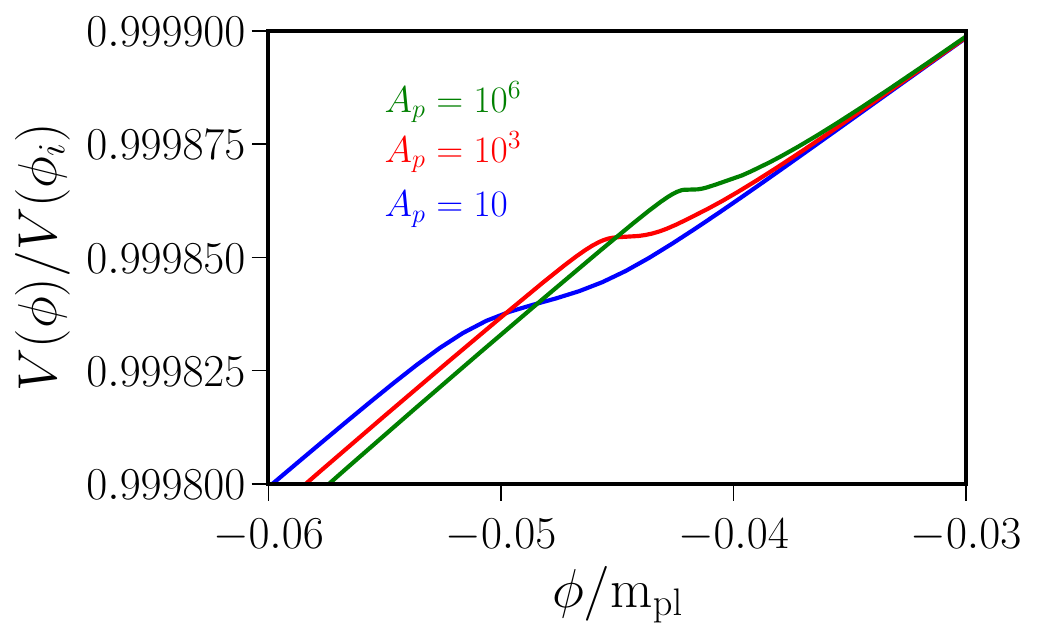}
\quad
\includegraphics[width = 0.437\textwidth]{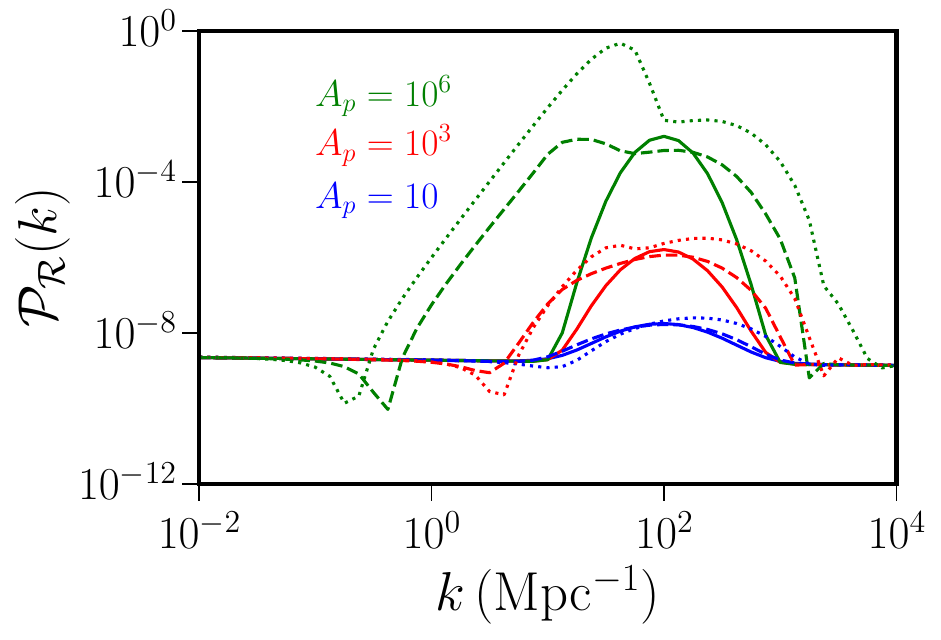}
\end{center}
\caption{(Left panel) Reconstructed potential from the power spectrum with a peak given by \eq{eq:peakP} for different values of $A_p$. Here we have used $r=10^{-4}$. (Right panel) Power spectra calculated numerically from the reconstructed potentials (dashed lines). For comparison, we also present the spectra from the reconstructed potentials which are obtained using the standard slow-roll approximation (dotted lines), as well as the input spectra (solid lines).}
\label{fig:peakedP}
\end{figure}

In the left panel of Figure~\ref{fig:peakedP}, we show the reconstructed potential from the given power spectrum with different values of $A_p$ using the tensor-to-scalar ratio $r=10^{-4}$. It can be seen that a small plateau around $\phi/\mpl \sim -0.04$ is enough to produce a peak high enough to give $\calP_\calR(k_c) = \calO(0.01)$, which may lead to the production of PBHs at that scale. In the right panel of Figure~\ref{fig:peakedP}, we have also computed numerically the power spectrum from the reconstructed potential for each $A_p$ (dashed lines). For comparison, we also show the input power spectrum \eq{eq:peakP} (solid line) as well as those calculated from the reconstructed potential using the standard slow-roll approximation (dotted line).

As we may well have expected, we can reproduce the input power spectrum if the height of the peak is not too large. It is notable that GSR works very reliably around $\calO(10)$ variation of the amplitude of $\calP_\calR$. Meanwhile, as the height of the peak increases significantly, our reconstruction programme works less effectively. This is evident from the right panel of Figure~\ref{fig:peakedP} where the power spectra with $A_p = 10^3$ and $10^6$ deviate from the input. This is not surprising as the GSR expansion parameter $g$, facing the features in the potential, becomes very large: $|g| = \calO(10)$ or even bigger. Still, it is very interesting to note that the height and location of the peak are captured in the reconstructed spectra.


\section{Conclusions}
\label{sec:conc}

We have proposed a new method to reconstruct the inflaton potential within the picture of canonical single field inflation, directly from the given primordial power spectrum which may deviate significantly from near scale invariance. Our approach relies on a more generalized slow-roll approximation without assuming the constancy of the slow-roll parameters. Therefore this method is valid for the potential with features where the spectral index of the power spectrum may change with scales.

As a specific example other than featureless power-law spectrum, we have used the power spectrum with features given analytically by \eq{eq:featuredP} to reconstruct the corresponding potential with different values of the tensor-to-scalar ratio. 
We have confirmed that the reconstructed potential exhibits a smooth step, in agreement with the underlying model. As a consistency check, we have calculated numerically from the reconstructed potential the power spectrum, which shows a very good agreement with the input power spectrum as shown in Figure~\ref{fig:featuredP}. If we can fix the large-scale primordial power spectrum with smaller error in the near future~\cite{Hazra:2016fkm,Yoshiura:2020soa}, our method presented in this article would provide a reliable way of determining the inflaton potential directly from power spectrum. The observations of the tensor power spectrum would eliminate the ambiguity in determining the integration constant in our program.

We have also applied our method to the power spectrum with a high peak that may lead to the formation of PBHs. If the peak is as high as $10^6$, the validity of GSR is broken since the expansion parameter of GSR becomes bigger than unity. In this case, the power spectrum calculated from the reconstructed potential deviates from the input power spectrum. However, it is very interesting that still the reconstructed potential captures the qualitative information about the height and location of the peak in the given power spectrum. Thus our method can, within limited validity, be useful even for the power spectrum that leads to the formation of PBHs.

\subsection*{Acknowledgements}

We thank Suro Kim, Tomo Takahashi, Masahide Yamaguchi and Chulmoon Yoo for helpful comments and discussions.
This work is supported in part by the National Research Foundation of Korea Grant 2019R1A2B5B01070181 (KYC and RNR) and 2019R1A2C2085023 (JG). 
SK is supported by the Korea Initiative for fostering University of Research and Innovation Program of the National Research Foundation of Korea (2020M3H1A1077095).
KYC and JG also acknowledge the Korea-Japan Basic Scientific Cooperation Program supported by the National Research Foundation of Korea and the Japan Society for the Promotion of Science (2020K2A9A2A08000097).
JG is further supported in part by the Ewha Womans University Research Grant of 2020 (1-2020-1630-001-1). 
JG is grateful to the Asia Pacific Center for Theoretical Physics for hospitality while this work was under progress.

\end{document}